\documentclass[10pt]{article}

\usepackage{amsmath}
\usepackage{amsfonts}
\usepackage{amssymb}
\usepackage{amsthm}
\usepackage{stmaryrd} 
\usepackage{braket}
\usepackage{empheq}
\usepackage{graphicx}
\usepackage{subfigure}
\usepackage{hyperref}
\usepackage{footmisc}
\usepackage{appendix}
\usepackage{epsf}
\usepackage{epsfig}
\usepackage{tikz,tikz-3dplot}
\usetikzlibrary{decorations.markings,decorations.pathmorphing}
\usepackage{caption}
\usepackage{pgfplots}
\usepackage[utf8]{inputenc}
\usepackage[english]{babel}
\usepackage[T1]{fontenc}
\usepackage{sectsty}
\usepackage{authblk}

\subsubsectionfont{\normalfont\itshape}
\usepackage{fullpage}

\renewcommand{\bar}{\overline}
\renewcommand{\leq}{\leqslant}

\newcommand{\bbone}{{\text{\usefont{U}{bbold}{m}{n}\char49}}}

\DeclareMathOperator{\tr}{tr}

\theoremstyle{plain}
\newtheorem{theorem}{Theorem}[section]
\newtheorem{proposition}[theorem]{Proposition}
\newtheorem{corollary}[theorem]{Corollary}

\theoremstyle{definition}

\newtheorem{remark}[theorem]{Remark}

\title{A proof that no-signalling implies microcausality \\ in quantum field theory}
\author[a b]{Antoine Soulas \thanks{\texttt{antoine.soulas@univie.ac.at} ; ORCID: 0000-0003-0952-6060} }
\affil[a]{\small{Faculty of Physics, University of Vienna, Boltzmanngasse 5, 1090 Vienna (Austria)}}
\affil[b]{\small{IQOQI Vienna, Austrian Academy of Sciences, Boltzmanngasse 3, 1090 Vienna (Austria) }}
\date{}

\begin{document}
\maketitle

\abstract{We study some logical interrelationships between fundamental properties in (relativistic) quantum theories. An operational no-signalling condition is first introduced in the context of quantum mechanics, where we prove its equivalence to an apparently weaker version restricted to ideal measurements, and to a property of factorization of the evolution unitary operator. We then translate this condition in quantum field theory and prove that it implies both microcausality and the spin-statistics theorem, in the ideal case of pointwise measurements implemented in the projection postulate sense. This provides an argument (often invoked but apparently missing in the literature) to see microcausality as a necessary condition for the compatibility of spacelike separated operations.}


\section*{Acknowledgements}
I would like to thank Dimitri Petritis for the great freedom he has granted me during my PhD. I also thank Laurent Bruneau, Dmitry Chernyak, Guilherme Franzmann and Augustin Vanrietvelde for careful rereading, as well as Časlav Brukner, Maria Eftychia Papageorgiou, Nicolás Medina Sánchez and Robin Simmons for stimulating discussions and sharp remarks.

\section*{Statements and Declarations}
\textbf{Funding:} the research leading to these results received funding from Quantum Science Austria (QuantA). \\
\textbf{Competing Interests:} the author has no competing interests to declare that are relevant to the content of this article.\\
\textbf{Ethics approval, Consent, Data, Materials and code availability, Authors’ contribution statements:} not applicable. 

\newpage
\section{Introduction} \label{intro}

Microcausality counts among the basic requirements that ensure the consistency of quantum field theories (QFT). It is in particular present in the Wightman \cite{streater2000pct} and the Haag-Kastler axioms \cite{haag1964algebraic}. As we will see, standard QFT textbooks generally introduce microcausality as an assumption, vaguely motivated by invoking the compatibility of spacelike measurements without more details, despite it being a crucial ingredient to prove the spin-statistics theorem. More generally, most of them don't check the compatibility between the quantum formalism and the disappearance of the notion of instantaneity in special relativity.

Even in the literature on the foundations of QFT, an elementary derivation of microcausality from the relativistic constraints seemed to be missing. Authors generally proceed the other way round, justifying the axiom as a \textit{sufficient} condition to guarantee some desired properties, typically involving properties like no-signalling, separability, independence of operations, tensor product structure \cite{summers1990independence, redei2010operational,  ruetsche2011interpreting, halvorson2006algebraic}. After reviewing the different approaches to motivate the microcausality axiom, Wright concludes that:

\begin{quotation}
‘... the axiom of microcausality is not well motivated on its own. What is clear is that supporters of microcausality must do more than gesture towards no-signalling results or the use of the tensor product structure as motivation for postulating the axiom and dig deeper to find a more stable, rigorous and justified foundation to stand the axiom on. There is still room for a good motivation to be found’. \cite{wright2012quantum}
\end{quotation}

The aim of this paper is to present an original (to the author's knowledge) derivation of microcausality from an operational no-signalling condition, thereby justifying microcausality as a \textit{necessary} condition, in the restrictive case where measurements are pointwise and implemented as a simple projection postulate. Here is a list of the different properties that we will study in the sequel. First, in the context of quantum mechanics (QM):

\paragraph{(C)} \textit{For all quantum systems composed of two entangled subsystems, any physical operation on the first must leave invariant the statistics of the second, if the two are spacelike separated.} \\

\paragraph{(MC)} \textit{For all quantum systems composed of two entangled subsystems, any non-selective measurement of the first must leave invariant the statistics of the second, if the two are spacelike separated.} \\

\paragraph{(F)} \textit{The unitary evolution operator of the bipartite system takes the factorized form $U = U_1 \otimes U_2$.} \\

Then, specific to the QFT case:

\paragraph{(M)} \textit{For all quantum fields $\Phi$ and spacelike intervals $x-y$, we have $[\Phi(x) , \Phi^\dagger(y)]_{\pm} = 0$ where $[ \;  , \; ]_{\pm}$ stands for an anti-commutator or a commutator depending on the fermionic or bosonic nature of $\Phi$.} \\

\paragraph{(S)} \textit{Scalar and vector fields correspond to bosons, while Dirac fields correspond to fermions.} \\

The paper is organized as follows. We begin with a historical review of the treatment of these topics and a recap on some of the state-of-the-art knowledge in Section \ref{review}. The no-signalling condition (C) is then motivated and translated mathematically in the context of QM in Section \ref{condition (C)}, where we also study some of its main properties. We then move to QFT in Section \ref{microcausality}, and look for its implications concerning the microcausality hypothesis. Precisely, we prove in this paper the following results:
\begin{itemize}
\item (F), (C) and (MC) are logically equivalent in finite dimensional QM;
\item in the case of pointwise and ideal measurements, a QFT version of (MC) implies both microcausality (M) and the spin-statistic theorem (S).
\end{itemize}

To be clear, we don't claim to end the debate on the microcausality axiom in QFT, our argument being based on a rather simplistic notion of measurement; as we will see, more realistic situations are still currently being investigated. Our aim is mainly to provide an elementary but important proof that was apparently missing.

\section{Previous works} \label{review}
There now exists a rich literature on this crucial topic related to many other fundamental questions. The reader must be aware that finding appropriate definitions that capture the interesting properties was — and still remains — an important part of the challenge, making it harder to compare different studies that do not always share the same definitions. 

In the context of QM, an early work of Lüders (1951) \cite{luders1951zustandsanderung, luders2006concerning} determined the form that the measurement projection postulate should take for general density matrices, and investigated the compatibility of two such measurements. He was able to prove the logical equivalence between (MC) (equation (16) in the reference) and the commutativity of the two measured observables. In particular, his derivation made no reference to a spacetime nor to a tensor product structure of the Hilbert space. However, still unsatisfied with the presumed instantaneity of the wavefunction collapse, Bloch (1967) \cite{bloch1967some} pointed out some remaining inconsistencies of quantum theory when considered within a relativistic spacetime, notably: ‘it appears that either causality or Lorentz covariance of wavefunctions must be sacrificed’. In reply, Hellwig and Kraus (1970) \cite{hellwig1970formal} clarified Bloch's ideas, checked a simple version (MC) (equation (6)) based on an assumption called ‘locality’, namely the commutation of projectors associated to spacelike measurements. They compare their treatment to the more complicated one of Schlieder (1968) \cite{schlieder1968einige} who had recently proved the logical equivalence between (MC) and the commutation of spacelike separated observables. Much later, Busch and Singh (1998) \cite{busch1998luders} reconsidered Lüder's initial derivation and extended his result to certain special pairs of unsharp observables. This sparkled an interest in trying to characterize abstractly the set of operations that obey no-signalling, prior to any tensor product factorization. These works are reviewed by Beck \cite[Chapter 3]{beck2021local}; in fact: ‘it turned out that it is not easy to answer this question in full generality (\dots) it is indeed possible to mathematically construct formal sets of state transformers so that given effects obey no signalling with respect these state transformers but at the same time do not obey the related commutativity conditions’. While all these works are partly relevant for QFT in that they help understanding the intuition behind the microcausality axiom, they are restricted to QM. In particular, they can only establish a commutation relation for quantum observables, but by no means (anti)commutation relations of quantum fields.

Another important class of results in QM are the various ‘no-communication theorems’, which generally involve the derivation of (MC) or (C) from the tensor product structure of a bipartite system, plus additional assumptions like (F). The first general proof of (F) implies (MC) was detailed by Ghirardi, Rimini and Weber (1980) in \cite{ghirardi1980general}. Many other proofs followed \cite{bussey1982super, jordan1983quantum, bell1985exchange, redhead1987incompleteness, scherer1993problem}, but Kennedy (1995) \cite{kennedy1995empirical} reviewed them and concluded that (i) these proofs are disguised versions of a single theorem (ii) they are actually circular, because no-signalling was tacitly assumed by von Neumann (and all subsequent formulations of QM) in the very construction of the tensor product structure of a composite system. D'Ariano (2007) \cite{d2007no} developed a general operational framework to further explore the links between no-signalling and other fundamental properties. He similarly noticed the central role played by the tensor product factorization, and found that the latter can not be entirely reconstructed from the commutativity of local transformations, unless one embraces the additional principle of ‘local observability’ guaranteeing that complete information of a composite system can be obtained by means of local experiments only.  Considering how deeply rooted in the foundations of QM is the no-signalling principle, some authors have rather studied the possibility of using (C) as an axiom of the theory. The seminal paper by Popescu and Rohrlich (1994) \cite{popescu1994quantum}, however, revealed that no-signalling alone was not sufficient to characterize the non-local correlations permitted by QM for entangled systems, but for example Simon, Bužek and Gisin (2001) \cite{simon2001no} argued that the linearity of QM can still be derived from the Born rule and the no-signalling condition.

With the advent of the XXIth century came the quantum information era. Bruss, D'Ariano, Macchiavello and Sacchi (2000) \cite{bruss2000approximate} provided the first formulation of the no-signalling theorem for general CPTP maps. Beckman, Gottesman, Nielsen and Preskill's article (2001) \cite{beckman2001causal} was a milestone: the two systems were now called Alice and Bob's parts, the evolution operators were renamed quantum operations, and a distinction between operations which are causal (not allowing communication) and localizable (realizable with local unitaries) was introduced. Although their treatment is restricted to QM, the authors' ultimate aim was to contribute to the long-standing problem of characterizing the set of observables in QFT, plagued by several issues such as Sorkin's impossible measurements \cite{sorkin1993impossible}, already pointed out by Dirac \cite{dirac1958principles}. The fact that (F) implies (C) was clearly stated from the beginning and the converse was also proved in finite dimension (Theorem 7). In the present paper, we will strengthen the latter, by showing in Theorem \ref{factorization_theorem} that the apparently weaker condition (MC) is actually sufficient to derive (F), thereby obtaining the equivalence between the PVM condition (MC) and the general channel version (C). Another notable article is \cite{schumacher2005locality}, in which Schumacher and Westmoreland (2005) exhibited three ways of expressing locality (Locality (III) being very similar to (C)), and showed their mathematical equivalence. Many works followed, and the condition (C) is now well-known in the quantum information literature (although the major reference \cite{nielsen2010quantum} does not mention it), especially for those working on causal decomposition \cite{lorenz2021causal} and quantum causal models \cite{barrett2021cyclic}.

In the context of QFT, things are significantly more involved. Although Haag and other founders of algebraic quantum field theory (AQFT) adopted an operational interpretation of the local algebras, it took a long time before the implementation of local measurements in (A)QFT were explored in depth \cite{fraser2023note}. A veritable ‘zoo’ of properties have been proposed (to quote the word of \cite{earman2014relativistic}), with complex logical interrelationships. Haag and Kastler (1964) \cite{haag1964algebraic} first hoped to derive the microcausality axiom from the physically more meaningful condition of ‘statistical independence’, but it turned out not to ensure (M) \cite{de1973commutativity}. Many other properties were formulated and studied in relation to (M), as reviewed by Summers (1990) \cite{summers1990independence}, Earman and Valente (2014) \cite{earman2014relativistic} and Calderón (2024) \cite{calderon2024causal}. Let us simply mention the property of ‘$\mathcal{A}-\mathcal{B}$-uncorrelatedness’ invented by Buchholz and Summers (2005) \cite{buchholz2005quantum}, which is experimentally testable in principle and indeed implies (M), as well as the concept of $C^*$-separability, close to (C), introduced by Rédéi and Valente (2010) \cite{redei2010local} to better encapsulate the notion of no-signalling for general operations. Conversely, note that (M) does not imply (C) in general in AQFT. Indeed, it is shown in \cite{redei2010local} (Proposition 6) that ‘a no-signalling theorem does not hold for general operations defined on local algebras’ even in the presence of (M). The difficulty arises from the fact that the admissible operations on a region of spacetime are not easily expressed, in short because the von Neumann algebras are typically of type III. Some attempts to better delimitate the admissible no-signalling operations in AQFT can be found in \cite{van2021relativistic}.

Parallel to AQFT, another major approach to study signalling properties in QFT, in view of addressing Sorkin's problem, relies on detector models such as the Unruh-DeWitt detector. The characteristic feature of these models is the coupling of a non-relativistic quantum mechanical system, the detector, with a quantum field. The advantage of such a strategy is to provide a realistic and yet rigorous implementation of the measurement process in QFT, certainly more satisfying than the simplistic idealization that we will use in Section \ref{microcausality}. Cliche and Kempf (2010) \cite{cliche2010relativistic} employed information theoretic techniques to explicitly compute the classical and quantum capacities of the channel induced between two \textit{pointwise} Unruh-DeWitt detectors. They showed that, assuming microcausality, ‘both the classical and the quantum channel capacities are strictly zero when [the two] are spacelike separated’, hence ‘causality [(C)] in the channel follows directly from microcausality [(M)]’. Martín-Martínez (2015) \cite{martin2015causality} showed that (M) guarantees no-signalling (although defined quite differently from (C)) between two \textit{compactly supported} detectors at second order in perturbation theory. A non-perturbative argument was given by de Ramón, Papageorgiou and Martín-Martínez (2021) in \cite{de2021relativistic}, based on an assumption of causal factorization  (namely the factorization of the joint scattering map, somehow a QFT version of (F)) of the detector-field interactions. The same authors further remarked (2023) \cite{de2023causality} that (M) is not sufficient to ensure (C) in detector models if noncompactly supported detectors are allowed, but under suitable assumptions one can still ‘define an effective light cone, outside of which signalling is negligible’.

Importantly, none of the above works provide a proof in the context of QFT that, under suitable assumptions, (C) implies (M). As we will see in the beginning of Section \ref{microcausality}, such a derivation is nowhere to be found in the main QFT textbooks. In \cite[Section 4.2]{earman2014relativistic}, Earman and Valente claim that (M) is logically equivalent to (C) when measurement are implemented in the projection postulate sense, referring to the abovementionned article by Schlieder \cite{schlieder1968einige}. However, we argue that Schlieder's theorem applies to QM observables, not to quantum fields (not to mention the fact that the article is rather unclear, written in terms of quite old-fashionned concepts, and published in German). The closest paper we could find to our present aims is the one of Eberhard and Ross (1989) \cite{eberhard1989quantum}, in which the authors prove that (M) implies (C) in QFT when measurements are implemented following a projection postulate. Our goal is precisely to show the converse in Theorem \ref{theorem microcausality}, namely that (C) (actually (MC)) implies (M), at least in this simple description of measurements.

In several of the above references, it is acknowledged that, from a foundational point of view, it is hard to be fully satisfied with a no-signalling condition like (C), due to the numerous philosophical complexities entailed by the notion of causality. Bell (1990) \cite{bell2004speakable} asked what \textit{really} has to be constrained by the speed of light, and pointed out the anthropocentrist aspect of causality when merely defined as operational no-signalling. He also provided an original proof of (M) implies (C), computing the effects of a variation of a Lagrangian density on other observables. Other subtleties stemming from the delicate connection between causality and the no-signalling condition are discussed by Butterfield (2007) \cite{butterfield2007reconsidering}. To overcome the anthropocentrism of (C) and the ill-motivated aspect of (M), Beck and Lazarovici (2024) \cite{beck2024relativistic} have recently proposed a new property called ‘relativistic consistency’, which is necessary to obtain consistent statistical predictions across different Lorentz frames. This property is stronger than (C), but it has the advantage of always implying (M). 

The links between the notion of causality in QM (encoded in the tensor product structure which ensures (C)) and in QFT (guaranteed by (M)) were notably explored in \cite{di2023relativistic}, where the authors find that the latter implies the former ‘in a physically relevant approximation in the quantum dynamics of a massive scalar quantum field coupled to two localised systems’. 

The present review of the literature is clearly not exhaustive. For further references, see for instance \cite{fraser2023note} for a detailed history of the treatment of local measurements in QFT, \cite{papageorgiou2024eliminating} for a recap of the different approaches to solve Sorkin's problem in (A)QFT, \cite{papageorgiou2023local} for further references on detector models, \cite{wright2012quantum} for a discussion on the existing arguments to motivate the microcausality axiom.

\section{The condition (C) in quantum mechanics} \label{condition (C)}
\subsection{Literal formulation} \label{literal}

There are two different sources of instantaneity in QM that could cause troubles when trying to build a relativistic quantum theory. When two subsystems are entangled, they must be considered as a whole\footnote{It has been experimentally confirmed that non-local correlations are satisfied (almost) immediately: the results of \cite{zbinden2001experimental} allow to ‘set a lower bound on the speed on quantum information to $10^7c$, \textit{i.e.} seven orders of magnitude larger than the speed of light.’}, therefore:
\begin{itemize}
\item any physical operation on the first instantaneously affects the whole state no matter how far the other part may be,
\item any measurement performed on the first allows the observer to apply a selective measurement or ‘collapse’ on the whole state, which is generally presented as an instantaneous update. 
\end{itemize}

However, only the first item is \textit{a priori} problematic, because the epistemic update of an observer having acquired new information has no reason to be constrained by the speed of light. As Bell wrote in \cite{bell2004speakable}: ‘When the Queen dies in London (may it long be delayed) the Prince of Wales, lecturing on modern architecture in Australia, becomes instantaneously King’, and he could have added: from the Queen and her entourage's point of view. On the other hand, what has to be constrained by special relativity are the physically predictable phenomena, so that no experiment conducted on the King can determine faster than light whether the Queen is alive. For this to be prevented in QM despite the non-locality of the entanglement phenomenon, the consistency condition (C) has to hold. We will also consider the more specific consistency condition (MC), dealing only with measurements, because it is easier to manipulate mathematically and will turn out to be equivalent to (C).

Such conditions are generally referred to as ‘(operational) no-signalling conditions’. Here, our choice of the letter (C) may stand for ‘consistency condition’, but also ‘covariance’ and ‘causality’. We consider this terminology slightly more accurate, because (i) strictly speaking, special relativity does not forbid anything to travel at superluminal speeds \cite[Chapter 3]{maudlin2011quantum}\cite{butterfield2007reconsidering}, only causal loops have to be rejected for the logical consistency of the theory (ii) covariance and causality are two separate notions. Indeed, if (C) were not satisfied (\textit{a fortiori} (MC)), the theory would face these two types of inconsistencies: 
	
\begin{itemize}
\item \textit{Non-covariance} — Consider two entangled quantum systems that violate the condition (C). Then there exists an experimental protocol concerning the second system that yields different statistical results depending on whether a certain operation has been performed on the first system or not, such that the two are spacelike separated. Thus one can find a reference frame in which the operation on the first system happens \textit{before} the measurement on the second, and another reference frame in which it happens \textit{after}. Consequently, the statistical predictions of the theory depend on the reference frame.
\item \textit{Causal paradoxes} — Consider a reference frame in which Alice and Bob are apart, motionless, and share $N$ entangled pairs that violate the condition (C) as well as two synchronized clocks. At $t=0$, Alice performs the suitable measurement on each of the $N$ subsystems if she wants to communicate the bit 0, or does nothing if she wants to communicate the bit 1. At $t=0^+$, Bob performs the corresponding measurement on each of the subsystems in his possession: the statistical distribution he obtains allows him to distinguish whether Alice has sent the bit 0 or 1, with an error margin arbitrarily small when $N$ goes larger. It is then well known that if such faster-than-light communications were possible, one could produce causal loops like the grandfather paradox.
\end{itemize}

Alternatively, if (C) holds, an entangled pair cannot be used to convey information. It also ensures that the theory is consistent with an instantaneous effect of physical interactions (including measurements) on the total state, even though the notion of instantaneity is frame dependent. Different observers may write different states for the entangled pair, but they agree on the statistics.

\subsection{Mathematical formulation} \label{mathematical}
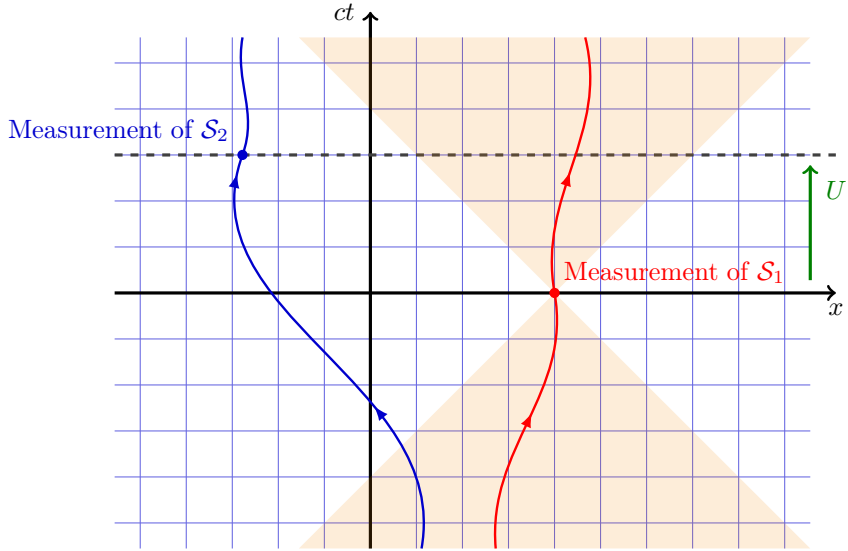
\begin{figure}[h]
\centering
\begin{tikzpicture}[scale=1.7]
\colorlet{myblue}{blue!80!black}
\colorlet{mypurple}{blue!40!red!80!black}
\colorlet{myorange}{orange!40!yellow!95!black}
  \def\xmax{2}
  \def\ext{4*\d} % extension
  \def\xmaxp{2.2} % maximum of rotated axis
  \def\Nlines{5} % number of world lines (at constant x/t)
  \pgfmathsetmacro\d{0.9*\xmax/\Nlines} % grid size
  \pgfmathsetmacro\ang{atan(1/3)} % angle between x and x' axes
  \coordinate (O) at (0,0);
  \coordinate (A) at (-1,1.08);
  \coordinate (X) at (\xmax+\ext+0.2,0);
  \coordinate (T) at (0,\xmax+0.2);
  \coordinate (C) at (45:\xmaxp+0.2);
  \coordinate (E) at (4*\d,0);
  %Grid
  \message{Making world lines...^^J}
  \foreach \i [evaluate={\x=\i*\d;}] in {1,...,\Nlines}{
    \message{  Running i/N=\i/\Nlines, x=\x...^^J}
    \draw[myblue!60,line width=0.4]   (-\x,-\xmax) -- (-\x,\xmax);
    \draw[myblue!60,line width=0.4]   ( \x,-\xmax) -- ( \x,\xmax);
    \draw[myblue!60,line width=0.4] (-\xmax,-\x) -- (\xmax+\ext,-\x);
    \draw[myblue!60,line width=0.4] (-\xmax, \x) -- (\xmax+\ext, \x);}
  \foreach \i [evaluate={\x=(\Nlines+\i)*\d;}] in {1,...,4}{
    \message{Running i/N=\i/\Nlines, x=\x...^^J}
    \draw[myblue!60,line width=0.4] (\x,-\xmax) -- ( \x,\xmax);}
  %Axes
  \draw[->,  very thick] (0,-\xmax) -- (T) node[left=2] {$ct$};
  \draw[->, very thick] (-\xmax,0) -- (X) node[below=0] {$x$};
  \draw[very thick, dashed, darkgray] (-\xmax,1.08) -- (\xmax+\ext+0.2,1.08) ;
   \draw[->, very thick, green!50!black] (\xmax+\ext, 0.1) -- (\xmax+\ext, 1) node[below right=2] {$U$};
  %Cone
  \begin{scope}[shift={(E)}]
    \fill[myorange!70!red,opacity=0.15]
      (0,0) -- (\xmax,\xmax) -- (-\xmax,\xmax) -- (\xmax,-\xmax) -- (-\xmax,-\xmax) -- cycle;
  \end{scope}
  %Particles worldlines
  \draw[myblue, line width=0.9, decoration={markings,mark=at position 0.27 with {\arrow{latex}},
                                      mark=at position 0.76 with {\arrow{latex}}},postaction={decorate}]
      (0.2*\xmax,-\xmax) to[out=80,in=-110] (A) to[out=70,in=-100] (-0.5*\xmax,\xmax);
  \draw[red, line width=0.9, decoration={markings,mark=at position 0.27 with {\arrow{latex}},
                                            mark=at position 0.74 with {\arrow{latex}}},postaction={decorate}]
      (0.49*\xmax,-\xmax) to[out=95,in=-80] (E) to[out=100,in=-76] (0.84*\xmax,\xmax);
      
  \fill[myblue] (A) circle(0.04) node[above left=1.5] {Measurement of $\mathcal{S}_2$} ;
  \fill[red] (E) circle(0.04) node[above right] {Measurement of $\mathcal{S}_1$};
\end{tikzpicture}
\caption{\bf{Spacelike measurements}}
\label{spacelike measurements}
\end{figure}

The condition (C) can be given a precise mathematical formulation. Let $\mathcal{S}_1 + \mathcal{S}_2$ be two entangled systems described by a Hilbert space $\mathcal{H}_1 \otimes  \mathcal{H}_2$, prepared in a state $\rho$. The most general physical evolution that $\mathcal{S}_1$ may undergo is a channel $\rho \rightsquigarrow \Psi \otimes \bbone_2(\rho)$ for some CPTP map $\Psi$ acting only in $\mathcal{H}_1$. In this paper, we will suppose for simplicity that the pair $\mathcal{S}_1 + \mathcal{S}_2$ evolves after this interaction according to a unitary operator $U$ (stronger conditions could be formulated for non-isolated pairs, but at least this weaker condition must hold). Since the state of $\mathcal{S}_2$, obtained by tracing over $\mathcal{S}_1$, fully encodes the probabilities of any measurement on $\mathcal{S}_2$, the invariance of the statistics of $\mathcal{S}_2$ at a spacelike separated distance from the operation on $\mathcal{S}_1$ (as in Fig. \ref{spacelike measurements}) is expressed by: 

\paragraph{Mathematical formulation of (C)}
\[ \leqno{\textbf{(C)}}\hskip2cm \forall \rho, \forall \Psi, \forall U, \quad\quad \tr_1 \left( U \Psi \otimes \bbone_2(\rho) U^\dagger \right) =  \tr_1 \left( U \rho U^\dagger \right).  \] 

\textit{Or, equivalently, due to a property of the partial trace: }

\[ \forall \rho, \forall \Psi, \forall U, \forall \hat{B}, \forall y_0 \in  \mathrm{spec}{\hat{B}}, \quad\quad
\tr \left( U \Psi \otimes \bbone_2(\rho) U^\dagger (\bbone_1 \otimes \Pi^{(2)}_{y_0}) \right) =  \tr \left( U \rho U^\dagger (\bbone_1 \otimes \Pi^{(2)}_{y_0}) \right).
\] \\

To formulate (MC), let also $\hat{A}$ (resp.\ $\hat{B}$) be an observable of $\mathcal{H}_1$ (resp.\ $\mathcal{H}_2$), and write its spectral decomposition $\hat{A} = \sum_{x \in \mathrm{spec}(\hat{A})} x \Pi^{(1)}_{x}$ (resp.\ $\hat{B} = \sum_{y \in \mathrm{spec}(\hat{B})} y \Pi^{(2)}_{y}$) where the $\Pi^{(1)}_{x}$ (resp.\ $\Pi^{(2)}_{y}$) are the spectral projectors of the observable. If the operators are not compact, this remains possible in the rigged Hilbert space formalism \cite{gadella2002unified}. When a non-selective measurement of $A$ is performed on $\mathcal{S}_1$, the whole state evolves to $ \sum_{x \in \mathrm{spec}(\hat{A})} (\Pi^{(1)}_{x} \otimes \bbone_2)  \rho (\Pi^{(1)}_{x} \otimes \bbone_2)$. After which $\mathcal{S}_1 + \mathcal{S}_2$ evolves, as previously, according to a unitary $U$ in $\mathcal{H}_1 \otimes \mathcal{H}_2$. For the theory to be consistent, whether the measurement of $\hat{A}$ has been performed or not must not modify the statistics of $\mathcal{S}_2$ at a spacelike separated distance (as in Fig. \ref{spacelike measurements}), hence the following condition:

\paragraph{Mathematical formulation of (MC)} 
   \[ \leqno{\textbf{(MC)}}\hskip2cm   \forall \rho, \forall U, \forall \hat{A}, \quad\quad \tr_1\left( \sum_{x \in \mathrm{spec}{\hat{A}}} U (\Pi^{(1)}_{x} \otimes \bbone_2) \rho (\Pi^{(1)}_{x}  \otimes \bbone_2) U^\dagger \right) =  \tr_1(U \rho U^\dagger). \]

\textit{Or, equivalently, due to a property of the partial trace: }

\begin{align}
\forall \rho, \forall U, \forall \hat{A}, & \forall \hat{B}, \forall y_0 \in  \mathrm{spec}{\hat{B}}, \nonumber \\
& \sum_{x \in \mathrm{spec}{\hat{A}}} \tr \left( U (\Pi^{(1)}_{x} \otimes \bbone_2) \rho (\Pi^{(1)}_{x} \otimes \bbone_2)  U^\dagger (\bbone_1 \otimes \Pi^{(2)}_{y_0}) \right) = \tr \big( U \rho U^\dagger (\bbone_1 \otimes \Pi^{(2)}_{y_0}) \big). \label{condition (MC) math}
\end{align}

Of course, a non-selective measurement $\rho \rightsquigarrow \sum_{x \in \mathrm{spec}(\hat{A})} (\Pi^{(1)}_{x} \otimes \bbone_2) \rho (\Pi^{(1)}_{x} \otimes \bbone_2)$ is a particular CPTP map, therefore \textbf{(C) implies (MC)}. As noticed in \cite{beckman2001causal}, it is equivalent by linearity to only require (C) or (MC) to hold for initial states of the form $\rho = \rho_1 \otimes \rho_2$.

\begin{remark}
\textbf{On the unavoidable probabilistic nature of fundamental physics.} Conditions (C) and (MC) only constrain the \textit{statistics}. This is because special relativity only imposes the covariance of what is predictable by a given theory; and QM, being a probabilistic theory, only predicts the statistics. Similarly, the ability to transmit information depends on the best theory available to the communicators. If they can't do better than probabilistic predictions, Alice has to be able to modify the statistical results of a repeated experiment on Bob's side in order to send him a bit. 

Alternatively, if they had a deterministic (in particular: hidden-variable) theory superseding QM, only one run could suffice. Such a theory should then be constrained by a stronger version of (C) where the word ‘statistics’ is replaced by ‘particular outcomes’. But this is precisely forbidden by Bell's theorem \cite{bell1964einstein} which implies that such a theory would necessarily display non-local features. More precisely, either the outcome or parameter independence assumption would have to be violated \cite{sep-bell-theorem}, allowing in any case for superluminal signalling and covariance issues. See also \cite{helling2019no} for a simple analytic example of non-locality in Bohmian mechanics compared to the standard QM treatment, where it is made clear that no probabilistic knowledge other than the one given by the wavefunction is compatible with special relativity, and \cite{valentini2002signal} for a general proof in hidden-variable theories. Quantum entanglement seemed spooky to Einstein because of the conviction that nothing in the physical universe could in principle be unpredictable, \textit{i.e.} inaccessible to human physics,\footnote{This metaphysical stance is exemplified in his famous ‘God does not play dice’; indeed, note how positivist this remark is: it is not that God necessarily plays dice, but only that the human mind may not be able to access better than statistically the way God plays.} but it is actually the contrary: in a world where entanglement exists, QM is the only non-spooky theory! This remark proves that we will never be able to build a deterministic theory supplanting QM (at least not without a radical change in our physicists' paradigms). This is also why the terms added to the Schrödinger equation in collapse models will always be stochastic, ‘because otherwise [they] would allow for faster-than-light communication’ \cite{bassi2013models} and why Bohmian mechanics, despite being a hidden-variable theory reproducing QM's results, does not add to it any predictive power.
\end{remark}

\subsection{The factorization property (F)}
Before moving to QFT, let's first derive some important properties of the conditions (C) and (MC). Keeping the notations introduced above, the factorization property (F) reads:

\paragraph{(F)} \textit{The unitary evolution operator of $\mathcal{S}_1 + \mathcal{S}_2$ takes the factorized form $U = U_1 \otimes U_2$.} \\

\begin{proposition}[\textbf{No-communication theorem}] \label{(F) => (C)} 
(F) implies (C).
\end{proposition}

\begin{proof}
It is enough by linearity to check the condition (C) on states of the form $\rho = \rho_1 \otimes \rho_2$. Assuming (F), we may write:

\begin{align*}
\tr \left( U \Psi \otimes \bbone_2( \rho) U^\dagger \; (\bbone_1 \otimes \Pi^{(2)}_{y_0}) \right) &=  \tr \left(  \Psi (\rho_1) \otimes \rho_2 \; (\bbone_1 \otimes U_2^\dagger \Pi^{(2)}_{y_0} U_2 ) \right) \\
&= \tr\Big( \Psi (\rho_1)\Big) \tr\left( \rho_2 U_2^\dagger \Pi^{(2)}_{y_0} U_2 \right) \\
&= \tr( \rho_1) \tr\left( \rho_2 U_2^\dagger \Pi^{(2)}_{y_0} U_2 \right) \\
&= \tr \left(  \rho_1 \otimes \rho_2 (\bbone_1 \otimes U_2^\dagger \Pi^{(2)}_{y_0} U_2 ) \right) \\
&= \tr \left( U  \rho U^\dagger (\bbone_1 \otimes \Pi^{(2)}_{y_0}) \right) 
\end{align*}
where we have used the fact that $\Psi$ is a trace preserving map.
\end{proof}

\begin{remark}  
Rigorously speaking, the above computation suffices to exclude causal inconsistencies, but lacks an additional argument to ensure covariance. See the Annex \ref{covariance} for the details.
\end{remark}  

We need our theories to satisfy (C). This is usually ensured by the tensor product structure $\mathcal{H}_1 \otimes \mathcal{H}_2$ and by postulating (F) for isolated systems (\textit{a fortiori} for spacelike separated ones), in view of the no-communication theorem above. However, unlike (F), (C) is a consistency condition directly required by relativistic considerations, so it seems after all more natural to postulate the latter. One can then try to determine the set of unitaries compatible with (C). It turns out that the \textit{a priori} weaker condition (MC) actually suffices to obtain (F) for spacelike separated systems, as shown in the following Theorem. A weaker version of this result, namely that (C) implies (F), was already proved in \cite[Theorem 7]{beckman2001causal} using a completely different argument, also restricted to the finite dimensional case.

\begin{theorem}[$\mathbf{(MC) \Rightarrow (F)}$] \label{factorization_theorem}
If $\mathcal{S}_1$ and $\mathcal{S}_2$ are finite-dimensional, then (MC) implies (F).
\end{theorem}

\begin{proof}
Let's denote $\mathcal{H}_1$ and $\mathcal{H}_2$ the Hilbert spaces associated with $\mathcal{S}_1$ and $\mathcal{S}_2$, with $n_1 = \mathrm{dim}(\mathcal{H}_1)$ and $n_2=\mathrm{dim}(\mathcal{H}_2)$ finite. One can write $U$ in the following generic form: 

\begin{equation} 
U = \sum_{\substack{i \\ 1 \leq k,l \leq n_2}} \alpha_{ikl} T_i \otimes \ket{k} \bra{l} \label{decomposition}
\end{equation}
with $( \ket{k} \bra{l})_{1 \leq k,l \leq n_2}$ the canonical basis of $\mathcal{L}(\mathcal{H}_2)$ associated with an orthonormal basis $(\ket{k})_{1 \leq k \leq n_2} $, and $(T_i)_i$ a basis of $\mathcal{L}(\mathcal{H}_1)$. When replacing $U$ in the expression \eqref{condition (MC) math} for the condition (MC), one gets for all Hermitian operators $\hat{A}$ and $\hat{B}$ of $\mathcal{H}_1$ and $\mathcal{H}_2$ and for all $y_0 \in \mathrm{spec}(\hat{B})$: 

\begin{align*} 
& \forall \rho, \quad \tr \Bigg( \rho \; \sum_{\substack{i,k,l \\ j,k',l'}} \Big[ \sum_{x \in \mathrm{spec}(\hat{A})}  \Pi^{(1)}_{x} T_j^\dagger T_i \Pi^{(1)}_{x} - T_j^\dagger T_i \Big] \otimes  \alpha_{ikl} \bar{\alpha_{jk'l'}}  \ket{l'} \bra{k'} \Pi^{(2)}_{y_0}  \ket{k} \bra{l} \Bigg) =0 \\
\Rightarrow \quad & \sum_{\substack{i,k,l \\ j,k',l'}} \Big[ \sum_{x \in \mathrm{spec}(\hat{A})}   \Pi^{(1)}_{x} T_j^\dagger T_i \Pi^{(1)}_{x} - T_j^\dagger T_i   \Big] \otimes  \alpha_{ikl} \bar{\alpha_{jk'l'}}  \ket{l'} \bra{k'} \Pi^{(2)}_{y_0}  \ket{k} \bra{l} = 0.
\end{align*}

Note that $\hat{B}$ may be chosen arbitrarily, and in particular $\Pi^{(2)}_{y_0}$. For any pair $\{ k_1, k_2\} \subset \llbracket 1,n_2 \rrbracket$ and $\mu, \nu \in \mathbb{C}$ such that $\lvert \mu \rvert^2 + \lvert \nu \rvert^2 = 1$, one can define $\Pi^{(2)}_{y_0}$ to be the projector on the vector $\mu \ket{k_1} + \nu \ket{k_2}$, that is $ \lvert \mu \rvert^2  \ket{k_1} \bra{k_1} + \mu \bar{\nu} \ket{k_1} \bra{k_2} + \bar{\mu} \nu \ket{k_2} \bra{k_1} + \lvert \nu \rvert^2  \ket{k_2} \bra{k_2} $. Inserting into the previous equation divides it in four sums: 

\begin{align*}
\forall k_1, k_2, \quad \sum_{\substack{i,l \\ j,l'}} & \Big[ \sum_{x \in \mathrm{spec}(\hat{A})}   \Pi^{(1)}_{x} T_j^\dagger T_i \Pi^{(1)}_{x} - T_j^\dagger T_i   \Big] \\
& \otimes  \Big[ \lvert \mu \rvert^2  \alpha_{ik_1 l} \bar{\alpha_{jk_1l'}} + \mu \bar{\nu} \alpha_{ik_2 l} \bar{\alpha_{jk_1l'}} + \bar{\mu} \nu \alpha_{ik_1 l} \bar{\alpha_{jk_2 l'}} + \lvert \nu \rvert^2  \alpha_{ik_2 l} \bar{\alpha_{jk_2 l'}}   \Big]  \ket{l'} \bra{l} = 0.
\end{align*}
The particular cases $\mu = 1, \nu = 0$ or $\mu = 0, \nu = 1$ imply that the first and fourth terms actually always vanish. Setting $\mu = \nu = \frac{1}{\sqrt{2}}$ or $\mu = \frac{1}{\sqrt{2}}, \nu = \frac{i}{\sqrt{2}}$ leads to: 
\[ \left\{ \begin{array}{ll}
\displaystyle \sum_{\substack{i,l \\ j,l'}} & \Big[ \displaystyle \sum_{x \in \mathrm{spec}(\hat{A})}   \Pi^{(1)}_{x} T_j^\dagger T_i \Pi^{(1)}_{x} - T_j^\dagger T_i   \Big]  \otimes  \Big[ \frac{1}{2} \alpha_{ik_2 l} \bar{\alpha_{jk_1l'}} +  \frac{1}{2} \alpha_{ik_1 l} \bar{\alpha_{jk_2 l'}}  \Big] \ket{l'} \bra{l} = 0 \\
\displaystyle \sum_{\substack{i,l \\ j,l'}} & \Big[ \displaystyle \sum_{x \in \mathrm{spec}(\hat{A})}   \Pi^{(1)}_{x} T_j^\dagger T_i \Pi^{(1)}_{x} - T_j^\dagger T_i   \Big]  \otimes  \Big[ -\frac{i}{2} \alpha_{ik_2 l} \bar{\alpha_{jk_1l'}} +  \frac{i}{2} \alpha_{ik_1 l} \bar{\alpha_{jk_2 l'}}  \Big] \ket{l'} \bra{l} = 0
\end{array} \right. \]
and taking appropriate linear combinations of these shows that the second and third terms vanish as well. Therefore: 

\begin{align*}
\forall k,k', \quad & \sum_{\substack{i,l \\ j,l'}} \; \alpha_{ik l} \bar{\alpha_{jk' l'}} \Big[ \sum_{x \in \mathrm{spec}(\hat{A})}   \Pi^{(1)}_{x} T_j^\dagger T_i \Pi^{(1)}_{x} - T_j^\dagger T_i   \Big]  \otimes  \ket{l'} \bra{l} = 0 \\
\Rightarrow \quad \forall k,k',l,l', \quad & \sum_{i, j} \; \alpha_{ik l} \bar{\alpha_{jk' l'}} \Big[ \sum_{x \in \mathrm{spec}(\hat{A})}   \Pi^{(1)}_{x} T_j^\dagger T_i \Pi^{(1)}_{x} - T_j^\dagger T_i   \Big] = 0
\end{align*}
because $(\ket{l'} \bra{l})_{1 \leq l,l' \leq n_2}$ is a basis of $\mathcal{L}(\mathcal{H}_2)$. This being true for all Hermitian operators $\hat{A}$, the operator $\sum_{i,j} \alpha_{ikl} \bar{\alpha_{jk'l'}} \; T_j^\dagger T_i = (\sum_{i} \alpha_{ik'l'} T_i )^\dagger (\sum_{i} \alpha_{ikl} T_i)$ is diagonal in every orthonormal bases of $\mathcal{H}_1$, so it is a dilation: 
\[ \forall k, k',l,l', \exists \lambda_{kk'll'} \in \mathbb{C}:  \quad (\sum_{i} \alpha_{ik'l'} T_i )^\dagger (\sum_{i} \alpha_{ikl} T_i) = \lambda_{kk'll'} \bbone. \]

Now, there exist $k_0$ and $l_0$ such that $\lambda_{k_0 k_0 l_0 l_0} \neq 0$. Otherwise, for all $k$ and $l$ we would have $\sum_{i} \alpha_{ikl} T_i = 0$ (since $\mathcal{L}(\mathcal{H}_2)$ is a $C^*$-algebra satisfying $\lVert X^\dagger X \rVert = \lVert X \rVert^2$ for all $X \in \mathcal{L}(\mathcal{H}_2)$), and by linear independence of the $(T_i)_i$ this would imply that all the $\alpha_{ikl}$ vanish \textit{i.e.} $U=0$, which is not possible.

At present, pick any pair of indices $k,l$. 
\begin{itemize}
\item If $\lambda_{k k_0 l l_0} \neq 0$, by unicity of the inverse in finite dimension, we have $\frac{1}{\lambda_{k k_0 l l_0}} \sum_{i} \alpha_{ikl} T_i = \frac{1}{\lambda_{k_0 k_0 l_0 l_0}} \sum_{i} \alpha_{i k_0 l_0} T_i$ and since the $(T_i)_i$ are linearly independent: $\alpha_{ikl} = \frac{\lambda_{k k_0 l l_0}}{\lambda_{k_0 k_0 l_0 l_0}} \alpha_{i k_0 l_0}$. 
\item If $\lambda_{k k_0 l l_0} = 0$, since $\sum_{i} \alpha_{i k_0 l_0} T_i$ is invertible, we deduce that $\sum_{i} \alpha_{ikl} T_i = 0$, hence $\alpha_{ikl}=0$ for all $i$.
\end{itemize}
In any case:
\[\forall k,l, \exists \beta_{kl} \in \mathbb{C}: \forall i,  \quad  \alpha_{ikl} = \beta_{kl} \; \alpha_{i k_0 l_0}. \]

It is now possible to factorize: 
\[ U = \sum_{\substack{i \\ 1 \leq k,l \leq n_2}} \beta_{kl} \alpha_{ik_0l_0} T_i \otimes \ket{k} \bra{l} =  \left( \sum_{i}  \alpha_{ik_0l_0} T_i \right) \otimes \left( \sum_{1 \leq k,l \leq n_2} \beta_{kl} \ket{k} \bra{l} \right) = U_1 \otimes U_2 \]
where one can identify $U_1$ and $U_2$ with the evolution operators of $\mathcal{S}_1$ and $\mathcal{S}_2$, which are necessarily unitary since $U = U_1 \otimes U_2$ is. 
\end{proof}

This proof does not straightforwardly extends to the infinite dimensional case, because then $\mathcal{L}(\mathcal{H}_1)$ and $\mathcal{L}(\mathcal{H}_2)$ are not separable so they don't admit a countable basis; moreover, having merely $T^\dagger T$ does not guarantee the invertibility of $T$ anymore.

An immediate consequence of this Theorem is that (F), (C) and (MC) imply themselves circularly, therefore \textbf{(F), (C) and (MC) are logically equivalent} in finite dimensional QM. Another corollary is the following: while (C) and (MC) do not seem \textit{a priori} to be symmetrical with respect to $\mathcal{S}_1$ and $\mathcal{S}_2$, their equivalence to (F) implies such a symmetry. Said differently, $\mathcal{S}_1$ does not allow to communicate to $\mathcal{S}_2$ if and only if the converse is true.

\section{Microcausality} \label{microcausality}
Let's now move to QFT. Usually, standard QFT textbooks justify the microcausality hypothesis (M) by invoking, for once, the concept of measurement \cite[p.28]{peskin2018introduction} \cite[p.106]{itzykson2012quantum} \cite[p.121]{zee2010quantum}, but they generally make do with the affirmation that two spacelike measurements must be independent, without more explanations. Not only is the argument too vague, but it is hard to see immediately how the relation could differ according to the fermionic or bosonic nature of the field. Weinberg, on the other hand, makes an interesting remark: 

\begin{quotation}
‘The condition [(M)] is often described as a causality condition, because if $x - y$ is spacelike then no signal can reach $y$ from $x$, so that a
measurement of $\Phi$ at point $x$ should not be able to interfere with a measurement of $\Phi$ or $\Phi^\dagger$ at point $y$. Such considerations of causality are plausible for the electromagnetic field, any one of whose components may be measured at a given spacetime point, as shown in a classic paper of Bohr and Rosenfeld \cite{bohr1933frage}. However, we will be dealing here with fields like the Dirac field of the electron that do not seem in any sense measurable. The point of view taken here is that [(M)] is needed for the Lorentz invariance of the S-matrix, without any ancillary assumptions about measurability or causality.’  \cite[p.198]{weinberg1995quantum}
\end{quotation}
As Weinberg notes himself, microcausality is only a sufficient condition for the invariance of the $S$-matrix: 
\begin{quotation}
‘Theories of this class [satisfying (M)] are not the only ones that are Lorentz invariant, but the most general Lorentz invariant theories are not very different. In particular, there is always a commutation condition something like [(M)] that needs to be satisfied. This condition has no counterpart for non-relativistic systems, for which time-ordering is always Galilean-invariant. \textit{It is this condition that makes the combination of Lorentz invariance and QM so restrictive}.’ \cite[p.145]{weinberg1995quantum}
\end{quotation}

However, Weinberg's argument strongly relies on the use of normal-ordered fields. Arguably, this writing is only a computation convenience without physical meaning. Its main purpose is to get rid of the infinite constants that appear in certain computations, by making finite all the matrix elements of the operators manipulated. This operation is justified when the divergences have no influence in the considered context. Alternatively, the prediction of the Casimir effect or the Lamb shift by the vacuum energy is only possible \textit{without} normal-ordering (one substantially uses the fact that $\braket{0 \vert \Phi^2 \vert 0} \neq 0$ for a non-ordered $\Phi$, see \cite[p.111]{itzykson2012quantum}). It seems that a stronger argument is needed to justify the microcausality hypothesis (M). 

Note that (M) is especially crucial to prove the famous spin-statistics theorem (S). Here is how most QFT textbooks proceed to establish (S). Since we don't know the result yet, we have to compute both commutators and anti-commutators. A short calculation first shows that when $\Phi$ is a Dirac field, we have: 
\begin{equation} [\Phi(x) , \Phi^\dagger(y)]_{\pm} = \Delta_+(x-y) \mp \Delta_+(y-x) \label{commut_dirac} \end{equation}
depending on $\Phi$  describing fermions or bosons, \textit{i.e.} $[a(\vec{p}),a^\dagger(\vec{p'})]_{\pm}=\delta^{(3)}(\vec{p} - \vec{p'})$, and where $\Delta_+(x) = \int \frac{\mathrm{d}^3p}{(2\pi)^3 2p_0} e^{ip_\mu x^\mu}$ is shown to be a Lorenz invariant quantity.
Likewise, when $\Phi$ is a scalar or a vector field: 
\begin{equation} [\Phi(x) , \Phi^\dagger(y)]_{\pm} = \Delta_+(x-y) \pm \Delta_+(y-x) \label{commut_scalaire} \end{equation}
depending on $\Phi$  describing fermions or bosons. It is quite easy to see that when $x-y$ is spacelike, $\Delta_+(x-y) - \Delta_+(y-x) = 0$, while $\Delta_+(x-y) + \Delta_+(y-x)$ is not identically zero \cite[p.202]{weinberg1995quantum}. All these remarks, in addition to (M), imply (S). Not the reverse: these computations generally don't justify (M) but only the fermionic or bosonic nature of a field, whereas one can intuitively feel that they contain much more information. 

Keeping in mind the above (anti-)commutators computations, we now present a simultaneous derivation of the microcausality hypothesis and the spin-statistics theorem, from a QFT version of (MC). Our proof also relies on the following assumptions:
\begin{enumerate}
\item measurements are taken pointwise and represented by an operation in the projection postulate sense;
\item a quantum field is either bosonic or fermionic;
\item there is a one-to-one correspondance between measurable quantities and Hermitian operators (although this assumption could be weakened, see Remark \ref{observability} below). \\
\end{enumerate}

\begin{theorem}[$\bold{(MC) \Rightarrow (M) \, and \, (S)}$] \label{theorem microcausality}
Let $\Phi$ be a quantum field. Then, under the above assumptions, for all spacelike intervals $x-y$, we have $[\Phi(x) , \Phi^\dagger(y)]_{\pm} = 0$ where $[ \;  , \; ]_{\pm}$ stands for an anti-commutator or a commutator depending on $\Phi$ being a Dirac field (and in that case it describes a fermion) or a scalar or vector field (and in that case it describes a boson).
\end{theorem} 

\begin{proof}
Let's first suppose that $\Phi$ is a measurable field, in other words $\Phi(x)$ is Hermitian for all $x$ . As stated by Weinberg in the previous quote, it is for example the case for the electromagnetic field that describes the photon. One can write its spectral decomposition $\Phi(x) = \sum_{\lambda \in \mathrm{spec} \; \Phi(x)} \lambda \Pi^{(x)}_{\lambda}$ (this holds in general in the rigged Hilbert space formalism \cite{gadella2002unified}). This expression is only distributionally defined, so rigorously one should work with smeared fields, however we will stick to this notation for the sake of clarity. Being a measurable field, $\Phi$ must be constrained by a relativistic consistency condition of the kind studied in Section \ref{condition (C)}. Contrary to the QM case, however, there is now only one Hilbert space, the Fock space $\mathcal{H}_{Fock}$, and the system's state is given by a density matrix $\rho \in \mathcal{S}(\mathcal{H}_{Fock})$. For all spacelike intervals $x-y$, a measurement of $\Phi(x)$ doesn't affect the statistics of $\Phi(y)$ if and only if the following condition, variant of (MC), is satisfied: 

\begin{align} \label{(C) QFT}
\forall \rho, \forall \mu \in \mathrm{spec} \; \Phi(y), \quad \tr \Bigg( \Big( \sum_{\lambda \in \mathrm{spec} \; \Phi(x)}  \Pi^{(x)}_{\lambda} \rho  \Pi^{(x)}_{\lambda} \Big)  \Pi^{(y)}_{\mu} \Bigg) = \tr ( \rho \Pi^{(y)}_{\mu} )
\end{align}
where we implicitly moved to a reference frame $\mathcal{R}$ in which $x^0 = y^0$, so as to avoid to introduce the (non-covariant) unitary evolution operators. It yields: 

\begin{align*} 
& \forall \rho, \forall \mu \in \mathrm{spec} \; \Phi(y), \quad \tr \Bigg( \rho \Big( \sum_{\lambda \in \mathrm{spec} \; \Phi(x)}  \Pi^{(x)}_{\lambda} \Pi^{(y)}_{\mu}  \Pi^{(x)}_{\lambda} - \Pi^{(y)}_{\mu} \Big) \Bigg) = 0  \\
\Rightarrow \quad & \forall \mu \in \mathrm{spec} \; \Phi(y), \quad  \sum_{\lambda \in \mathrm{spec} \; \Phi(x)}  \Pi^{(x)}_{\lambda} \Pi^{(y)}_{\mu}  \Pi^{(x)}_{\lambda}= \Pi^{(y)}_{\mu} \\
\Rightarrow \quad & \sum_{\lambda \in \mathrm{spec} \; \Phi(x)}  \Pi^{(x)}_{\lambda} \Phi(y) \Pi^{(x)}_{\lambda}= \Phi(y).
\end{align*}

Thus, $\Phi(y)$ is (block) diagonal in the eigenbasis of $\Phi(x)$, so they are codiagonalizable and $[\Phi(x) , \Phi(y)] = 0$. This relation \textit{a priori} holds in the frame $\mathcal{R}$, but when applying the appropriate representation of the Lorenz group under which $\Phi$ transforms, one sees that $\Phi(x)$ and $\Phi(y)$ commute in all reference frames. \\

If now $\Phi$ is not supposed to be Hermitian anymore, we still know that $\Phi \Phi^\dagger$ is. Applying what precedes to $\Phi \Phi^\dagger$ instead of $\Phi$, we obtain that for all $x-y$ spacelike, $[ \Phi(x) \Phi^\dagger(x) , \Phi(y) \Phi^\dagger(y) ] = 0$. Moreover, as $\Phi$ is not Hermitian, one shows as usual that for all $x$ and $y$, $[\Phi(x),\Phi(y)]=[\Phi^\dagger(x),\Phi^\dagger(y)]=0$ \footnote{This is because the only commutators that can appear between a creation and an annihilation operator are $[a(\vec{p}),a_c^{\dagger}(\vec{p'})]$ or $[a_c(\vec{p}),a^\dagger(\vec{p'})]$ (with the label $c$ standing for the antiparticle) which are zero if $a(\vec{p}) \neq a_c(\vec{p})$, \textit{i.e.} if the field is not Hermitian.}. But on the other hand, it is also possible to compute the commutator $[\Phi(x) \Phi^\dagger(x) , \Phi(y) \Phi^\dagger(y) ]$ using the commutation relations \eqref{commut_dirac} and \eqref{commut_scalaire}. Suppose for instance that $\Phi$ is a Dirac field. Since we don't know yet if it is a boson or a fermion, let's distinguish the possible cases: 

\begin{itemize}
\item if the particle described by $\Phi$ is a fermion, we have \eqref{commut_dirac} with the upper signs. Then: 

\begin{align*} 
\Phi(x) \underbrace{\Phi^\dagger(x) \Phi(y)}_{\substack{ = - \Phi(y) \Phi^\dagger(x) \\ - \Delta_+(x-y) \\ + \Delta_+(y-x)}} \Phi^\dagger(y) 
& = - \Phi(x) \Phi(y) \Phi^\dagger(x) \Phi^\dagger(y) + \Big(\Delta_+(y-x) - \Delta_+(x-y) \Big) \Phi(x) \Phi^\dagger(y) \\
& = - \Phi(y)  \underbrace{\Phi(x) \Phi^\dagger(y)}_{\substack{ = - \Phi^\dagger(y) \Phi(x) \\ + \Delta_+(x-y) \\ - \Delta_+(y-x)}} \Phi^\dagger(x) + \Big(\Delta_+(y-x) - \Delta_+(x-y) \Big) \Phi(x) \Phi^\dagger(y) \\
& =  \Phi(y) \Phi^\dagger(y) \Phi(x) \Phi^\dagger(x) \\ 
& \quad + \Big( \underbrace{\Delta_+(y-x) - \Delta_+(x-y)}_{=0 \text{ if } x-y \text{ spacelike}} \Big) (\Phi(y) \Phi^\dagger(x) + \Phi(x) \Phi^\dagger(y)),
\end{align*}

\item if the particle described by $\Phi$ is a boson, we have \eqref{commut_dirac} with the lower signs. Then 
\begin{align*} 
\Phi(x) \underbrace{\Phi^\dagger(x) \Phi(y)}_{\substack{ = \Phi(y) \Phi^\dagger(x) \\ - \Delta_+(x-y) \\ - \Delta_+(y-x)}} \Phi^\dagger(y) 
& = \Phi(x) \Phi(y) \Phi^\dagger(x) \Phi^\dagger(y) - \Big(\Delta_+(x-y) + \Delta_+(y-x) \Big) \Phi(x) \Phi^\dagger(y) \\
& = \Phi(y)  \underbrace{\Phi(x) \Phi^\dagger(y)}_{\substack{ = \Phi^\dagger(y) \Phi(x) \\ + \Delta_+(x-y) \\ + \Delta_+(y-x)}} \Phi^\dagger(x) - \Big(\Delta_+(x-y) + \Delta_+(y-x) \Big) \Phi(x) \Phi^\dagger(y) \\
& =  \Phi(y) \Phi^\dagger(y) \Phi(x) \Phi^\dagger(x) \\
& \quad + \Big(\underbrace{\Delta_+(x-y) + \Delta_+(y-x)}_{\substack{\text{non-identically zero } \text{if } x-y \text{ spacelike}}} \Big) (\Phi(y) \Phi^\dagger(x) - \Phi(x) \Phi^\dagger(y)).
\end{align*}
\end{itemize}

As a consequence, one recovers the commutation relation $[ \Phi(x) \Phi^\dagger(x) , \Phi(y) \Phi^\dagger(y) ] = 0$ imposed by the QFT translation of (MC) if, and only if, $\Phi$ is a fermionic field and in this case, the relation \eqref{commut_dirac} implies that $\{ \Phi(x) , \Phi^\dagger(y)\} = 0$ for all spacelike intervals $x-y$. These are indeed the statements (M) and (S). When $\Phi$ is a scalar or vector field, the proof is similar.
\end{proof}

\begin{remark} \label{observability}
This proof makes use, as usually in physics, of the identification between the notion of observable in the mathematical sense (Hermitian operator) and in the physical sense (quantity measurable by a concrete experimental protocol). However, the second is far more restrictive, since in practice we can only measure a few very specific observables. Rigorously speaking, only the physical notion of observable is constrained by (MC), since the latter must ensure the absence of inconsistencies between \textit{actual} measurements. Therefore, in the above proof, although $\Phi \Phi^\dagger$ is Hermitian, one could question the legitimacy of imposing it (MC). Of course, any mathematical observable could be in principle measured by applying a suitable unitary evolution to the system that would map its eigenbasis to the eigenbasis of a physically measurable observable. But is it satisfactory to rely on the idea that all unitary evolutions are \textit{a priori} feasible, even though we will never be able to implement them? Nonetheless, it is still possible to adapt the proof by replacing $\Phi \Phi^\dagger$ by an undoubtedly physical observable, such that a function of the components of $\hat{T}_{\mu \nu}$ or even the charge $\hat{Q}$, that one can express in terms of $\Phi$. For example, for a Dirac field, $\hat{T}_{\mu}^{\mu} = m \Phi^\dagger \gamma^0 \Phi$, would allow a quite similar proof.
\end{remark}

An immediate consequence of the above proof is the following. 

\begin{corollary}
A fermionic field is not measurable.
\end{corollary}

\begin{proof}
In the previous proof, we have seen that if a field is an observable (namely if $\Phi = \Phi^\dagger$), then the condition (MC) implies that for all $x-y$ spacelike, $[\Phi(x) , \Phi(y)] = 0$. But if $\Phi$ is a fermionic field, it also satisfies $\{ \Phi(x) , \Phi(y) \} = 0$. Adding these two relations yields for all $x-y$ spacelike, $\Phi(x) \Phi(y) = 0$, which is too strong a constraint for a quantum field. In particular, $\Phi(x) \Phi(y) \ket{0}$ is the state containing two localized particles at $x$ and $y$; in any case it is a non-zero vector of the Fock space.
\end{proof}

This result is not new, and can also be derived, for instance, by the fact that the equation $\Phi = \Phi^\dagger$ is neither covariant nor independent of the representation of the gamma matrices \cite{pal2011dirac}, so it can't be linked to a physically meaningful property such as being measurable. But it is interesting to see that it can be also derived from fundamental considerations about no-signalling. Again, our argument relies on the widespread identification between the notions of mathematical and physical observable. Here, $\Phi$ may indeed be Hermitian, but is it a physically measurable quantity subject to the condition (MC)?

\section{Conclusion} \label{conclusion}
In this paper, we have investigated some logical interrelationships between fundamental properties in QM and QFT. Our starting point was the apparent incompatibility between special relativity and two kinds of instantaneities that seem to appear when considering the entanglement phenomenon. By the time it was historically developed, quantum theory was not built to integrate special relativity, so that it is always surprising to contemplate their ‘peaceful coexistence’ (expression coined by Shimony in \cite{shimony1978metaphysical}).

We have first formulated a quantum mechanical no-signalling condition, ensuring the absence of covariance and causality issues. Its most general channel version (C) turned out to be equivalent (in finite dimension) to the apparently more restrictive PVM formulation (MC). When translated in the context of QFT, it allowed to see microcausality (M) as a necessary condition to respect this relativistic constraint, rather than a merely sufficient one, thereby motivating (M) as an axiom of QFT. Let's emphasize once more that, contrary to (F) or (M), (C) is not an arbitrary postulate but a consistency criterion that must be valid for any relativistic quantum theory: it ‘costs nothing’ to be assumed because it stems from the constraints of special relativity. Again, the proof of Theorem \ref{theorem microcausality} employs a very idealized implementation of the measurement, but the aim was mostly to carefully write down a simple argument that seemed absent in the existing literature.

Further works could include include the study of the logical interrelationships between (C), (MC) and (F) in the infinite dimensional QM case and in QFT. It would also be interesting to see how our proof could be adapted to more general statistics than bosons or fermions, such as Green's \cite{green1953generalized} or Wang's \cite{wang2025particle} parastatistics.

\begin{appendices}
\section{Condition (C) ensures covariance} \label{covariance}
In \S\ref{condition (C)}, we have motivated the importance of (C) by the fact that two kinds of inconsistencies appear if it were not satisfied. The proof of Proposition \ref{(F) => (C)} suffices to exclude the second kind (causality), but lacks an additional argument to get rid of the first kind (covariance). Indeed, it has assumed the choice of a fixed reference frame $\mathcal{R}$. To make things precise, let $t_1$ and $t_2$ be the instants in $\mathcal{R}$ corresponding to the operation undergone by $\mathcal{S}_1$ and the measurement performed on $\mathcal{S}_2$, respectively. In another reference frame, however, neither the temporal axis nor the unitary operators are conserved. Even the possible outcomes of the measurements may undergo a Lorenz transformation, if they correspond to the position or momentum observable for example. The invariance of the statistics between two reference frames $\mathcal{R}$ and $\mathcal{R}'$ may be written as follows (one assumes for instance $t_2 < t_1$ in $\mathcal{R}$ and denote $t_0$ an earlier time at which the two entangled systems were separated): 

\[ \left\{ \begin{array}{ll}
\bullet \; \tr_1 \left( U^{(\mathcal{R})}_{t_0,t_2} \rho U^{(\mathcal{R}) \dagger}_{t_0,t_2} \right) = \tr_1 \left( U^{(\mathcal{R'})}_{t'_0,t'_2} \rho U^{(\mathcal{R'}) \dagger}_{t'_0,t'_2} \right) \quad & \text{if }  t'_2 < t'_1. \\ \\

\bullet \; \tr_1 \left( U^{(\mathcal{R})}_{t_0,t_2} \rho U^{(\mathcal{R}) \dagger}_{t_0,t_2} \right)
 = \tr_1 \left( U^{(\mathcal{R'})}_{t'_1,t'_2} \;  \Psi \otimes \bbone_2 \left( U^{(\mathcal{R'})}_{t'_0,t'_1} \rho U^{(\mathcal{R'}) \dagger}_{t'_0,t'_1} \right)  U^{(\mathcal{R'}) \dagger}_{t'_1, t'_2} \right) \quad & \text{if } t'_2 > t'_1 
 \end{array} \right.  \] 
 
The first line is simply the covariance of the theory in the absence of measurements: we suppose it already established. The only case to examine involves the changes of reference frames that reverse the temporal order of $t_1$ and $t_2$. In particular, since the only potential discontinuity occur when the sign of $t_2-t_1$ flips, it suffices to write the second line above in the limit $\varepsilon, \varepsilon' \rightarrow 0$ of an infinitesimal change of frame with $t_2 = t_1 - \varepsilon$ and $t'_2 = t'_1 + \varepsilon'$. In that case, $U^{(\mathcal{R'})}_{t'_a,t'_b} \rightarrow  U^{(\mathcal{R})}_{t_a,t_b}$ for all instants $t_a$ and $t_b$. Moreover, if $y_0$ is a position or momentum variable, $y'_0 \rightarrow y_0$ and the two are related by a (bijective) Lorentz transformation that does not change the multiplicity of the eigenvalues, which ensures the continuity of the spectral projectors. The condition to verify reads: 
\[ \tr_1 \left( U^{(\mathcal{R})}_{t_0,t_2} \rho U^{(\mathcal{R}) \dagger}_{t_0,t_2} \right)
 = \tr_1 \left( U^{(\mathcal{R})}_{t_1,t_2} \;  \Psi \otimes \bbone_2 \left( U^{(\mathcal{R})}_{t_0,t_1} \rho U^{(\mathcal{R}) \dagger}_{t_0,t_1} \right)  U^{(\mathcal{R}) \dagger}_{t_1, t_2} \right) \]
but since $U^{(\mathcal{R})}_{t_0,t_2} = U^{(\mathcal{R})}_{t_1,t_2} U^{(\mathcal{R})}_{t_0,t_1}$, this is nothing but the condition (C) that has been proved applied to $U^{(\mathcal{R})}_{t_0,t_1} \rho U^{(\mathcal{R}) \dagger}_{t_0,t_1}$ instead of $\rho$.  
\end{appendices}

\bibliographystyle{siam}
\bibliography{Biblio_microcausality}

\begin{thebibliography}{10}

\bibitem{barrett2021cyclic}
{\sc J.~Barrett, R.~Lorenz, and O.~Oreshkov}, {\em Cyclic quantum causal
  models}, Nature communications, 12 (2021), p.~885.
\newblock doi : 10.1038/s41467-020-20456-x.

\bibitem{bassi2013models}
{\sc A.~Bassi, K.~Lochan, S.~Satin, T.~P. Singh, and H.~Ulbricht}, {\em Models
  of wave-function collapse, underlying theories, and experimental tests},
  Reviews of Modern Physics, 85 (2013), p.~471.
\newblock doi : 10.1103/RevModPhys.85.471.

\bibitem{beck2021local}
{\sc C.~Beck}, {\em Local quantum measurement and relativity}, Springer, 2021.

\bibitem{beck2024relativistic}
{\sc C.~Beck and D.~Lazarovici}, {\em Relativistic consistency of nonlocal
  quantum correlations}, Entropy, 26 (2024), p.~548.

\bibitem{beckman2001causal}
{\sc D.~Beckman, D.~Gottesman, M.~A. Nielsen, and J.~Preskill}, {\em Causal and
  localizable quantum operations}, Physical Review A, 64 (2001), p.~052309.
\newblock doi : 10.1103/PhysRevA.64.052309.

\bibitem{bell1964einstein}
{\sc J.~S. Bell}, {\em On the {E}instein {P}odolsky {R}osen paradox}, Physics
  Physique Fizika, 1 (1964), p.~195.
\newblock doi : 10.1103/PhysicsPhysiqueFizika.1.195.

\bibitem{bell2004speakable}
\leavevmode\vrule height 2pt depth -1.6pt width 23pt, {\em La nouvelle
  cuisine}, in Speakable and Unspeakable in Quantum Mechanics, Cambridge
  University Press, 2004, p.~234.
\newblock doi : 10.1017/CBO9780511815676.

\bibitem{bell1985exchange}
{\sc J.~S. Bell, A.~Shimony, M.~A. Horne, and J.~F. Clauser}, {\em An exchange
  on local beables}, Dialectica,  (1985), pp.~85--110.

\bibitem{bloch1967some}
{\sc I.~Bloch}, {\em Some relativistic oddities in the quantum theory of
  observation}, Physical Review, 156 (1967), p.~1377.
\newblock doi : 10.1103/PhysRev.156.1377.

\bibitem{bohr1933frage}
{\sc N.~Bohr and L.~Rosenfeld}, {\em Zur {F}rage der {M}essbarkeit der
  elektromagnetischen {F}eldgr{\"o}ssen}, Kgl. Danske Vidensk. Selskab.
  Math.-Fys. Medd, 12 (1933), p.~3.

\bibitem{bruss2000approximate}
{\sc D.~Bruss, G.~D’Ariano, C.~Macchiavello, and M.~Sacchi}, {\em Approximate
  quantum cloning and the impossibility of superluminal information transfer},
  Physical Review A, 62 (2000), p.~062302.

\bibitem{buchholz2005quantum}
{\sc D.~Buchholz and S.~J. Summers}, {\em Quantum statistics and locality},
  Physics Letters A, 337 (2005), pp.~17--21.

\bibitem{busch1998luders}
{\sc P.~Busch and J.~Singh}, {\em L{\"u}ders theorem for unsharp quantum
  measurements}, Physics Letters A, 249 (1998), pp.~10--12.

\bibitem{bussey1982super}
{\sc P.~Bussey}, {\em ‘{S}uper-luminal communication’ in
  {E}instein-{P}odolsky-{R}osen experiments}, Physics Letters A, 90 (1982),
  pp.~9--12.

\bibitem{butterfield2007reconsidering}
{\sc J.~Butterfield}, {\em Reconsidering relativistic causality}, International
  Studies in the Philosophy of Science, 21 (2007), pp.~295--328.
\newblock doi : 10.1080/02698590701589585.

\bibitem{calderon2024causal}
{\sc F.~Calder{\'o}n}, {\em The causal axioms of algebraic quantum field
  theory: A diagnostic}, Studies in History and Philosophy of Science, 104
  (2024), pp.~98--108.

\bibitem{cliche2010relativistic}
{\sc M.~Cliche and A.~Kempf}, {\em Relativistic quantum channel of
  communication through field quanta}, Physical Review A—Atomic, Molecular,
  and Optical Physics, 81 (2010), p.~012330.

\bibitem{d2007no}
{\sc G.~M. D'Ariano}, {\em No-signalling, dynamical independence and the local
  observability principle}, Journal of Physics A: Mathematical and Theoretical,
  40 (2007), p.~8137.

\bibitem{de1973commutativity}
{\sc B.~De~Facio and D.~C. Taylor}, {\em Commutativity and causal
  independence}, Physical Review D, 8 (1973), p.~2729.

\bibitem{de2021relativistic}
{\sc J.~de~Ram{\'o}n, M.~Papageorgiou, and E.~Mart{\'\i}n-Mart{\'\i}nez}, {\em
  Relativistic causality in particle detector models: Faster-than-light
  signaling and impossible measurements}, Physical Review D, 103 (2021),
  p.~085002.

\bibitem{de2023causality}
\leavevmode\vrule height 2pt depth -1.6pt width 23pt, {\em Causality and
  signalling in noncompact detector-field interactions}, Physical Review D, 108
  (2023), p.~045015.

\bibitem{di2023relativistic}
{\sc A.~Di~Biagio, R.~Howl, {\v{C}}.~Brukner, C.~Rovelli, and
  M.~Christodoulou}, {\em Relativistic locality can imply subsystem locality},
  arXiv preprint arXiv:2305.05645,  (2023).

\bibitem{dirac1958principles}
{\sc P.~A.~M. Dirac}, {\em The principles of quantum mechanics}, Oxford
  University Press, 1958.

\bibitem{earman2014relativistic}
{\sc J.~Earman and G.~Valente}, {\em Relativistic causality in algebraic
  quantum field theory}, International Studies in the Philosophy of Science, 28
  (2014), pp.~1--48.

\bibitem{eberhard1989quantum}
{\sc P.~H. Eberhard and R.~R. Ross}, {\em Quantum field theory cannot provide
  faster-than-light communication}, Foundations of Physics Letters, 2 (1989),
  pp.~127--149.

\bibitem{fraser2023note}
{\sc D.~Fraser and M.~Papageorgiou}, {\em Note on episodes in the history of
  modeling measurements in local spacetime regions using {QFT}}, The European
  Physical Journal H, 48 (2023), p.~14.

\bibitem{gadella2002unified}
{\sc M.~Gadella and F.~G{\'o}mez}, {\em A unified mathematical formalism for
  the {D}irac formulation of quantum mechanics}, Foundations of Physics, 32
  (2002), pp.~815--869.

\bibitem{ghirardi1980general}
{\sc G.-C. Ghirardi, A.~Rimini, and T.~Weber}, {\em A general argument against
  superluminal transmission through the quantum mechanical measurement
  process}, Lettere al Nuovo Cimento (1971-1985), 27 (1980), pp.~293--298.
\newblock doi : 10.1007/BF02817189.

\bibitem{green1953generalized}
{\sc H.~S. Green}, {\em A generalized method of field quantization}, Physical
  Review, 90 (1953), p.~270.

\bibitem{haag1964algebraic}
{\sc R.~Haag and D.~Kastler}, {\em An algebraic approach to quantum field
  theory}, Journal of Mathematical Physics, 5 (1964), pp.~848--861.

\bibitem{halvorson2006algebraic}
{\sc H.~Halvorson and M.~M{\"u}ger}, {\em Algebraic quantum field theory},
  arXiv preprint math-ph/0602036,  (2006).

\bibitem{helling2019no}
{\sc R.~C. Helling}, {\em No signalling and unknowable {B}ohmian particle
  positions}, arXiv preprint arXiv:1902.03752,  (2019).
\newblock doi : 10.48550/arXiv.1902.03752.

\bibitem{hellwig1970formal}
{\sc K.-E. Hellwig and K.~Kraus}, {\em Formal description of measurements in
  local quantum field theory}, Physical Review D, 1 (1970), p.~566.
\newblock doi : 10.1103/PhysRevD.1.566.

\bibitem{itzykson2012quantum}
{\sc C.~Itzykson and J.-B. Zuber}, {\em Quantum field theory}, Courier
  Corporation, 2012.
\newblock ISBN : 9780486134697.

\bibitem{jordan1983quantum}
{\sc T.~F. Jordan}, {\em Quantum correlations do not transmit signals}, Physics
  Letters A, 94 (1983), p.~264.

\bibitem{kennedy1995empirical}
{\sc J.~B. Kennedy}, {\em On the empirical foundations of the quantum
  no-signalling proofs}, Philosophy of Science, 62 (1995), pp.~543--560.

\bibitem{lorenz2021causal}
{\sc R.~Lorenz and J.~Barrett}, {\em Causal and compositional structure of
  unitary transformations}, Quantum, 5 (2021), p.~511.
\newblock doi : 10.22331/q-2021-07-28-511.

\bibitem{luders1951zustandsanderung}
{\sc G.~L{\"u}ders}, {\em {\"U}ber die {Z}ustands{\"a}nderung durch den
  {M}essprozess}, Annalen der Physik, 443 (1951), pp.~322--328.

\bibitem{luders2006concerning}
\leavevmode\vrule height 2pt depth -1.6pt width 23pt, {\em Concerning the
  state-change due to the measurement process}, Annalen der Physik, 518 (2006),
  pp.~663--670.

\bibitem{martin2015causality}
{\sc E.~Mart{\'\i}n-Mart{\'\i}nez}, {\em Causality issues of particle detector
  models in {QFT} and quantum optics}, Physical Review D, 92 (2015), p.~104019.

\bibitem{maudlin2011quantum}
{\sc T.~Maudlin}, {\em Quantum non-locality and relativity: Metaphysical
  intimations of modern physics}, John Wiley \& Sons, 2011.

\bibitem{sep-bell-theorem}
{\sc W.~Myrvold, M.~Genovese, and A.~Shimony}, {\em {Bell’s Theorem}}, in The
  {Stanford} Encyclopedia of Philosophy, E.~N. Zalta, ed., Metaphysics Research
  Lab, Stanford University, {F}all 2021~ed., 2021.

\bibitem{nielsen2010quantum}
{\sc M.~A. Nielsen and I.~L. Chuang}, {\em Quantum computation and quantum
  information}, Cambridge university press, 2010.
\newblock doi : 10.1017/CBO9780511976667.

\bibitem{pal2011dirac}
{\sc P.~B. Pal}, {\em Dirac, {M}ajorana, and {W}eyl fermions}, American Journal
  of Physics, 79 (2011), pp.~485--498.
\newblock doi : 10.1119/1.3549729.

\bibitem{papageorgiou2024eliminating}
{\sc M.~Papageorgiou and D.~Fraser}, {\em Eliminating the ‘impossible’:
  Recent progress on local measurement theory for quantum field theory},
  Foundations of Physics, 54 (2024), pp.~1--75.

\bibitem{papageorgiou2023local}
{\sc M.-E. Papageorgiou}, {\em Local measurements in relativistic quantum
  information: localization and signaling}, PhD thesis, University of Waterloo,
  2023.

\bibitem{peskin2018introduction}
{\sc M.~E. Peskin}, {\em An introduction to quantum field theory}, CRC press,
  2018.
\newblock doi : 10.1201/9780429503559.

\bibitem{popescu1994quantum}
{\sc S.~Popescu and D.~Rohrlich}, {\em Quantum nonlocality as an axiom},
  Foundations of Physics, 24 (1994), pp.~379--385.

\bibitem{redei2010operational}
{\sc M.~R{\'e}dei}, {\em Operational independence and operational separability
  in algebraic quantum mechanics}, Foundations of Physics, 40 (2010),
  pp.~1439--1449.

\bibitem{redei2010local}
{\sc M.~R{\'e}dei and G.~Valente}, {\em How local are local operations in local
  quantum field theory?}, Studies in History and Philosophy of Science Part B:
  Studies in History and Philosophy of Modern Physics, 41 (2010), pp.~346--353.

\bibitem{redhead1987incompleteness}
{\sc M.~Redhead}, {\em Incompleteness, non locality and realism. {A}
  prolegomenon to the philosophy of quantum mechanics}, Revue Philosophique de
  la France Et de l'Etranger, 180 (1987), pp.~712--713.

\bibitem{ruetsche2011interpreting}
{\sc L.~Ruetsche}, {\em Interpreting quantum theories}, Oxford University
  Press, 2011.

\bibitem{scherer1993problem}
{\sc H.~Scherer and P.~Busch}, {\em Problem of signal transmission via quantum
  correlations and {E}instein incompleteness in quantum mechanics}, Physical
  Review A, 47 (1993), p.~1647.

\bibitem{schlieder1968einige}
{\sc S.~Schlieder}, {\em Einige {B}emerkungen zur {Z}ustands{\"a}nderung von
  relativistischen quantenmechanischen {S}ystemen durch {M}essungen und zur
  {L}okalit{\"a}tsforderung}, Communications in Mathematical Physics, 7 (1968),
  pp.~305--331.

\bibitem{schumacher2005locality}
{\sc B.~Schumacher and M.~D. Westmoreland}, {\em Locality and information
  transfer in quantum operations}, Quantum Information Processing, 4 (2005),
  pp.~13--34.
\newblock doi : 10.1007/s11128-004-3193-y.

\bibitem{shimony1978metaphysical}
{\sc A.~Shimony}, {\em Metaphysical problems in the foundations of quantum
  mechanics}, International Philosophical Quarterly, 18 (1978), pp.~3--17.
\newblock doi : 10.5840/ipq19781818.

\bibitem{simon2001no}
{\sc C.~Simon, V.~Bu{\v{z}}ek, and N.~Gisin}, {\em No-signaling condition and
  quantum dynamics}, Physical Review Letters, 87 (2001), p.~170405.

\bibitem{sorkin1993impossible}
{\sc R.~D. Sorkin}, {\em Impossible measurements on quantum fields}, in
  Directions in general relativity: Proceedings of the 1993 International
  Symposium, Maryland, vol.~2, 1993, pp.~293--305.
\newblock doi : 10.1017/CBO9780511524653.

\bibitem{streater2000pct}
{\sc R.~F. Streater and A.~S. Wightman}, {\em PCT, spin and statistics, and all
  that}, vol.~30, Princeton University Press, 2000.

\bibitem{summers1990independence}
{\sc S.~J. Summers}, {\em On the independence of local algebras in quantum
  field theory}, Reviews in Mathematical Physics, 2 (1990), pp.~201--247.

\bibitem{valentini2002signal}
{\sc A.~Valentini}, {\em Signal-locality in hidden-variables theories}, Physics
  Letters A, 297 (2002), pp.~273--278.

\bibitem{van2021relativistic}
{\sc T.~van~der Lugt}, {\em Relativistic limits on quantum operations}, arXiv
  preprint arXiv:2108.05904,  (2021).

\bibitem{wang2025particle}
{\sc Z.~Wang and K.~R. Hazzard}, {\em Particle exchange statistics beyond
  fermions and bosons}, Nature, 637 (2025), pp.~314--318.

\bibitem{weinberg1995quantum}
{\sc S.~Weinberg}, {\em The quantum theory of fields}, vol.~2, Cambridge
  university press, 1995.
\newblock doi : 10.1017/CBO9781139644167.

\bibitem{wright2012quantum}
{\sc J.~Wright}, {\em Quantum field theory: Motivating the axiom of
  microcausality}, Master's thesis, University of Waterloo, 2012.

\bibitem{zbinden2001experimental}
{\sc H.~Zbinden, J.~Brendel, N.~Gisin, and W.~Tittel}, {\em Experimental test
  of nonlocal quantum correlation in relativistic configurations}, Physical
  Review A, 63 (2001), p.~022111.
\newblock doi : 10.1103/PhysRevA.63.022111.

\bibitem{zee2010quantum}
{\sc A.~Zee}, {\em Quantum field theory in a nutshell}, vol.~7, Princeton
  university press, 2010.
\newblock doi : 10.1088/0264-9381/28/8/089003.

\end{thebibliography}
\end{document}